\documentclass[12pt]{iopart}
\usepackage{citesort}

\usepackage{graphicx}
\usepackage[usenames]{color}
\expandafter\let\csname equation*\endcsname\relax
\expandafter\let\csname endequation*\endcsname\relax
\usepackage{amssymb,amsmath}
\usepackage[square,comma,compress,numbers]{natbib}
\newcommand{\eq}[1]{Eq.~(\ref{#1})}
\newcommand{\fig}[1]{Fig.~\ref{#1}}
\newcommand{\be}[1]{\begin{equation}\label{#1}}
\newcommand{\ee}{\end{equation}}

\usepackage{float}

\begin{document}

\title{Frustrated double and single ionization in a two-electron triatomic molecule H$^+_3$}

\author{A. Chen}
\address{Department of Physics and Astronomy, University College London, Gower Street, London WC1E 6BT, United Kingdom}
\author{C. Lazarou}
\address{Department of Physics and Astronomy, University College London, Gower Street, London WC1E 6BT, United Kingdom}
\author{H. Price}
\address{Department of Physics and Astronomy, University College London, Gower Street, London WC1E 6BT, United Kingdom}
\author{A. Emmanouilidou}
\address{Department of Physics and Astronomy, University College London, Gower Street, London WC1E 6BT, United Kingdom}

\begin{abstract}
Using a semi-classical model, we study  the formation of highly excited neutral fragments  during the fragmentation of $\mathrm{H_3^+}$, a two-electron triatomic molecule,  driven by an intense near-IR laser field.
To do so, we first formulate a microcanonical distribution for arbitrary one-electron triatomic molecules.
 We then study 
frustrated double and single ionization in strongly-driven $\mathrm{H_3^+}$ and compute the kinetic energy release of the nuclei for these two processes. Moreover, we investigate the dependence of frustrated ionization on the  strength  of the laser field as well as on the geometry of the initial molecular state. 
  \end{abstract}
\date{\today}

\maketitle
 
\section{Introduction}
In recent years, the highly nonlinear phenomena present in  molecules driven by intense near-infrared (near-IR) laser fields have attracted a lot of interest \cite{Review_dynamics_molecues}. One such phenomenon is the formation of highly excited fragments via  frustrated tunnel ionization \cite{FTI_exp_Nubbemeyer2008,FTI_exp_Manschwetus2009}.  Formation of highly excited fragments has been observed in the diatomic molecules H$_2$ \cite{FTI_exp_Manschwetus2009} and N$_2$ \cite{FTI_N2_Nubbemeyer2009}, the Ar dimer \cite{FTI_Ar_Ulrich2010}
and, recently, in the triatomic molecule D$^+_3$ \cite{H3+_exp_McKenna2009}.

For strongly-driven multi-center molecules, the study of multi-electron dynamics and its interplay with nuclear motion  poses a great challenge both for  theory and experiment alike. 
Tracing the dynamics of the electrons and the nuclei at the same time is currently beyond the capabilities of ab-initio quantum mechanical techniques.  Quantum techniques can currently address strongly-driven one electron triatomic molecules in two dimensions \cite{Lefebvretriatomic2014}.
This difficulty is tackled by classical models which are faster compared to  quantum techniques and  provide significant insights  into  the multi-electron dynamics and the interplay of electron-nuclear motion. 

In previous studies, we have presented a  three-dimensional (3D) semi-classical model to describe double ionization (DI) and frustrated double ionization (FDI) through Coulomb explosion. We have done so in the context of strongly-driven  H$_2$ \cite{semiclassical_agapi2012, Toolkit_Agapi2014}. Our 3D method has several assets, namely, it treats the motion of the electrons and the nuclei at the same time and fully addresses the Coulomb singularity \cite{semiclassical_agapi2012,Toolkit_Agapi2014}. 
Regarding the latter, the propagation  involves the global regularization scheme described in \cite{regularisation_Heggie1974} and a time-transformed leapfrog propagation technique \cite{Leapfrog_Mikkola2002} in conjunction with the Bulirsch-Stoer method \cite{Numerical_Recipes2007, differential_Stoer1966}. Another asset of our 3D technique 
is that it allows  for each of the two electrons to tunnel during propagation using the Wentzel-Kramers-Brilouin (WKB) approximation   \cite{semiclassical_agapi2012,Toolkit_Agapi2014}. This  is important in order to accurately describe   enhanced ionization  during the fragmentation of strongly-driven molecules 
\cite{Enhanced_Niikura2002,Enhanced_Bandrauk1995,enhanced_Seideman1995,enhanced_Villeneuve1996,enhanced_Dehghanian2010,enhanced_wu2012}.
Our results for H$_{2}$ were in good agreement with experimental results \cite{FTI_exp_Manschwetus2009}. 

Very recently, we have generalized our 3D model to describe the fragmentation of strongly-driven  two-electron triatomic molecules. Incorporating tunneling during propagation in our model allows for a better description of frustrated tunnel ionization  compared to the description provided by other classical models for strongly-driven triatomic molecules \cite{classical_D3+_Lotstedt2011,classical_Yamanouchi_2013}. In \cite{Agapi2016submitted}, we study the kinetic energy release of the nuclei for  DI and FDI for strongly-driven D$_3^{+}$  when the latter is fragmenting from the state that is created via the reaction $\mathrm{D_{2}+D_{2}^{+}\rightarrow D_{3}^{+}+D}$. Considering this initial state allows us to compare our results  for the kinetic energy release with experimental results \cite{D3+_exp_McKenna2012}. We find our results in good agreement with experiment.  

In this paper,  we present in detail a microcanonical distribution for arbitrary one-electron triatomic molecules. This distribution is one of the main features of our 3D technique and we employ it  to describe the initial state of the electron that is initially bound in our studies of strongly-driven triatomic molecules.  Currently, in the literature,  a microcanonical distribution is available only for diatomic \cite{micro_canonical_Olson1989} but not for triatomic molecules. Using our recently developed 3D semi-classical technique, we study frustrated ionization for  H$^+_3$ when it is strongly-driven  from its ground state. The inter-nuclear distance of the equilateral  configuration of H$_{3}^{+}$ in its ground state is 1.65 a.u. \cite{H3_GS_R_Schwartz1967}. This distance   is smaller from the inter-nuclear distance of the initial state we consider for our study of strongly-driven D$_{3}^{+}$ \cite{Agapi2016submitted} which varies from 2.04 a.u. to 2.92 a.u.. In this work, besides the formation of one highly excited neutral fragment with one electron escaping (FDI), a process which we explore for a different initial state in \cite{Agapi2016submitted},  we also study the formation of two highly excited neutral fragments with no electrons escaping, that is, frustrated single ionization (FSI).  In addition, we investigate  the dependence of  FDI and FSI on the intensity  of the laser field. For FDI we also study its dependence on the  geometry of  the initial molecular state. We do the latter  by comparing our results for the   driven diatomic H$_2$ with our results for the  driven triatomic  H$^+_3$.
The paper is structured as follows. In section 2 we formulate the microcanonical distribution for arbitrary one-electron triatomic molecules. In section 3.2, we present our  results for FDI and FSI, while in section 3.3 we investigate  whether a larger number of  highly excited fragments is formed  in H$^+_3$ versus H$_2$.


\section{Microcanonical distribution for one-electron triatomic molecules}
In this section, we  formulate a one-electron microcanonical distribution for triatomic molecules. We denote the positions of the nuclei by $\mathrm{{\bf R}_a=(0,0,-R_{a b}/2)}$, $\mathrm{{\bf R}_{b}=(0,0,R_{a b}/2)}$ and $\mathrm{{\bf R}_c=(x_{c},0,z_{c})}$ and the inter-nuclear distances by $\mathrm{R_{a b}}$, $\mathrm{R_{a c}}$ and $\mathrm{R_{b c}}$, see \fig{fig_1}. One can show that the coordinates of the nucleus C are expressed in terms of the inter-nuclear distances as follows:
\begin{figure} [ht]
\centering
 \includegraphics[clip,width=0.3\textwidth]{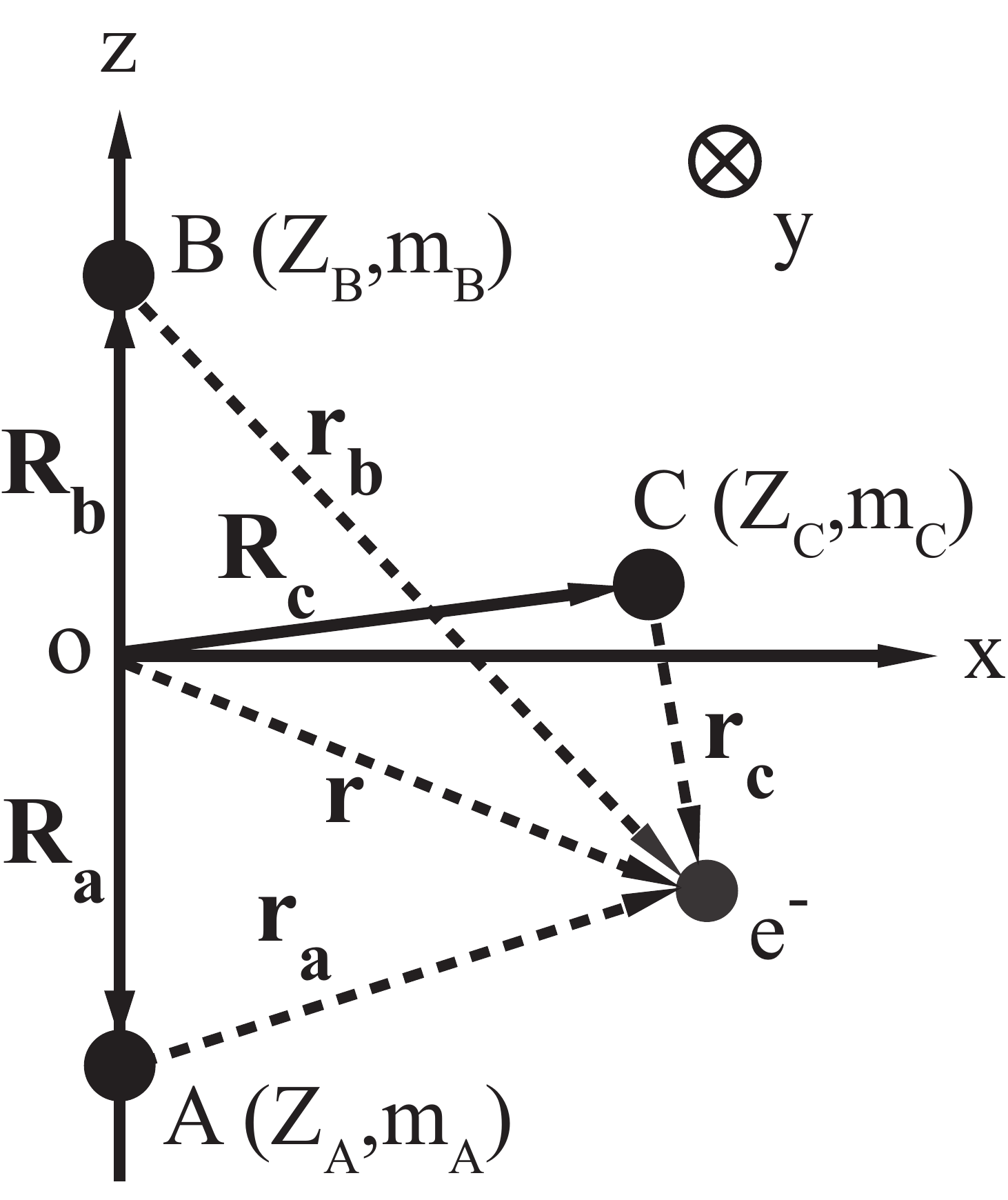}
\caption{The configuration of the triatomic molecule we use to set-up  the microcanonical distribution.}
\label{fig_1}
\end{figure}
\begin{equation}
\mathrm{z_{c}=\frac{R_{a c}^2-R_{bc}^2}{2 R_{a b}}}, \hspace{6pt}\mathrm{x_{c}=\pm\sqrt{R_{a c}^2-\left(\frac{R_{a c}^2-R_{bc}^2+R_{a b}^2}{2 R_{a b}}\right)^2}}.
\end{equation} 
 We denote the position vector of the electron by $\mathrm{{\bf r}}$ and the distances of the electron from the nuclei A, B and C  by $\mathrm{r_{a}=|{\bf r}-{\bf R}_a|}$,  $\mathrm{r_{b}=|{\bf r}-{\bf R}_b|}$ and $\mathrm{r_{c}=|{\bf r}-{\bf R}_c|}$, respectively. We then define the confocal elliptical coordinates $\mathrm{\lambda}$ and $\mathrm{\mu}$ using  the nuclei A and B as the foci of the ellipse, that is, 
 \begin{equation}
 \mathrm{\lambda=\frac{r_{a}+r_b}{R_{a b}}},  \hspace{6pt}\mathrm{\mu=\frac{r_{a}-r_b}{R_{a b}}},
 \end{equation}
 where $\mathrm{\lambda \in [1,\infty)}$ and $\mathrm{\mu \in [-1,1]}$. The third coordinate  $\mathrm{\phi \in [0,2 \pi]}$ is the angle between the projection of the position vector $\mathrm{{\bf r}}$ on the xy plane and the positive x axis; it thus defines the rotation angle around the axis that goes through the nuclei A and B.
 The potential of the electron  in the presence of the nuclei A, B and C, which have  charges $\mathrm{Z_{A}}$, $\mathrm{Z_{B}}$ and $\mathrm{Z_{C}}$, respectively, is given by 
 \begin{equation}
\mathrm{W(r_{a},r_{b},r_c)=-\frac{Z_A}{r_a}-\frac{Z_B}{r_b}-\frac{Z_C}{r_c}}.
\end{equation}
This potential is expressed   in terms of the confocal elliptical coordinates as follows

{\small
\setlength{\mathindent}{0cm}
\begin{eqnarray}
\mathrm{W(\lambda,\mu, \phi)}&&\mathrm{=-\frac{2}{R_{a b}}\left[\frac{Z_A}{\lambda+\mu} +\frac{Z_B}{\lambda-\mu} +\right.}\nonumber\\
&&\mathrm{\left.Z_{C}\left ( (\lambda^2+\mu^2-1)-\frac{4z_c}{R_{a b}}\lambda \mu-\frac{4x_{c}}{R_{a b}}\cos(\phi)\sqrt{(\lambda^2-1)(1-\mu^2)}+\frac{4(x_c^2+z_{c}^2)}{R_{a b}^2}\right)^{-\frac{1}{2}}\right ]}.
\end{eqnarray}
}

The one-electron microcanonical distribution is given by
\begin{equation}
\mathrm{f({\bf r}, {\bf p})}\propto \mathrm{\delta(E_{i}-\frac{p^2}{2}-W)},
\end{equation}
where $\mathrm{E_{i}=-I_p}$ is the ionization energy of the one-electron triatomic molecule. Note that the energy is given by $\mathrm{E=p^2/2+W}$. The electron momentum in terms of the confocal elliptical coordinates is expressed as follows
\begin{eqnarray}
\mathrm{p_x}&=&\mathrm{\sqrt{2(E-W(\lambda,\mu, \phi))}\cos(\phi_{p})\sqrt{1-\nu_{p}^2}}, \nonumber\\
\mathrm{p_y}&=&\mathrm{\sqrt{2(E-W(\lambda,\mu, \phi))}\sin(\phi_{p})\sqrt{1-\nu_{p}^2}}, \\
\mathrm{p_z}&=&\mathrm{\sqrt{2(E-W(\lambda,\mu, \phi))}\nu_p},\nonumber
\end{eqnarray}
where $\mathrm{\phi_{p}\in [0,2\pi]}$ and $\mathrm{\nu_{p}\in[-1,1]}$ define the momentum $\mathrm{\bf p}$ in spherical coordinates.
Transforming from $\mathrm{({\bf r}, {\bf p})\rightarrow(\lambda,\mu,\phi,E,\nu_{p},\phi_{p})}$ and integrating $\mathrm{f(\lambda,\mu,\phi,E,\nu_{p},\phi_{p})}$
over $\mathrm{E}\in(-\infty,0)$, $\mathrm{\phi_{p}}$ and $\mathrm{\nu_{p}}$ we find 
\begin{equation}
\mathrm{\rho(\lambda,\mu,\phi)\propto (\lambda^2-\mu^2)\sqrt{2(E_{i}-W(\lambda,\mu,\phi))}}.
\end{equation}

The  $\mathrm{\rho}$ distribution goes to zero and it is thus well-behaved when the electron is placed on top of either nucleus A or B. However, when $\mathrm{{\bf r}\rightarrow {\bf R}_{c}}$, i.e., the electron is placed on top
of nucleus C, $\mathrm{\rho(\lambda,\mu,\phi)\rightarrow \infty}$. We eliminate this singularity by introducing an additional transformation. Setting $\mathrm{\lambda=\lambda_{c}=(R_{ac}+R_{bc})/R_{ab}}$, $\mathrm{\phi=0}$ and expanding $\mathrm{\rho(\lambda_{c},\mu,0)}$ around $\mathrm{\mu=\mu_c=(R_{ac}-R_{bc})/R_{ab}}$ we find 
\begin{equation}
\mathrm{\rho(\lambda_c,\mu,0)\propto \frac{1}{|\mu-\mu_c|^{1/2}}},
\label{eq:sing}
\end{equation}
where $\mathrm{\lambda_c}$ and $\mathrm{\mu_c}$ are  the values of $\mathrm{\lambda}$ and $\mathrm{\mu}$, respectively, when the electron is placed on top of nucleus C. To eliminate the singularity in \eq{eq:sing}, we introduce a new variable $\mathrm{t}$, such that $\mathrm{t^{\gamma}=\mu-\mu_c}$. The new distribution takes the form

{\small
\setlength{\mathindent}{0cm}
\begin{eqnarray}
\tilde{\rho}(\lambda,t,\phi)&\propto&\left\{
\begin{array}{lll}
|t^{\gamma-1}|(\lambda^2-(t^{\gamma}+\mu_{c})^2)\sqrt{P(\lambda,t,\phi)} &\mathrm{for} &P(\lambda,t,\phi)\geq 0\\
& & \nonumber\\
0&\mathrm{for}  &P(\lambda,t,\phi)<0,
\end{array}
\right.\\
&&\\
\mathrm{P(\lambda,t,\phi)}&=&\mathrm{2E_{i}+\frac{4}{R_{a b}}}\mathrm{\left[\frac{Z_A}{\lambda+t^{\gamma}+\mu_c} +\frac{Z_B}{\lambda-t^{\gamma}-\mu_c}+Z_{C}\left((\lambda^2+(t^{\gamma}+\mu_c)^2-1)-\frac{4z_c}{R_{a b}}\lambda (t^{\gamma}+\mu_c)-\right.\right.}\nonumber\\
&&\mathrm{\left.\left.\frac{4x_{c}}{R_{a b}}\cos(\phi)\sqrt{(\lambda^2-1)(1-(t^{\gamma}+\mu_c)^2)}+\frac{4(x_c^2+z_{c}^2)}{R_{ab}^2}\right)^{-\frac{1}{2}}\right]}.\nonumber
\end{eqnarray}
}
Since $\mathrm{\mu \in [-1,1]}$, $\mathrm{t^{\gamma}}$ and $\mathrm{t}$ take both negative and positive values and therefore, if we choose one  $\mathrm{\gamma}$ for all values of $\mathrm{\mu}$, $\mathrm{\gamma}$ must be odd.   Moreover, to avoid the singularity when the electron is placed on top of nucleus C, $\mathrm{\gamma}$ must be such that  $\mathrm{t^{\gamma-1}/t^{\gamma/2}\rightarrow 0}$, i.e., $\mathrm{\gamma \geq 2}$. Combining the above two conditions, yields $\mathrm{\gamma=3,5,7,...}$. The new distribution 
$\mathrm{\tilde{\rho}(\lambda,t,\phi)}$ goes to zero when the electron is placed on top of nucleus C, i.e., when $\mathrm{\lambda=\lambda_c}$, $\mathrm{t=0}$ and $\mathrm{\phi=0,2\pi}$. 

To set up the initial conditions we find $\mathrm{\lambda_{max}}$ so that $\mathrm{p^2/2=E_{i}-W}>0$ and equivalently 
 $\mathrm{P(\lambda,t,\phi)\geq 0}$. We then find the maximum value $\mathrm{\tilde{\rho}_{max}}$ of the distribution $\mathrm{\tilde{\rho}(\lambda,t,\phi)}$, for the allowed values  of the parameters $\mathrm{\lambda}$, $\mathrm{t}$ and $\mathrm{\phi}$. We next generate the uniform random numbers $\mathrm{\lambda \in [1, \lambda_{max}]}$, $\mathrm{t \in [t_{min},t_{max}]}$, $\mathrm{\phi \in [0, 2\pi]}$ and $\mathrm{\chi \in [0, \tilde{\rho}_{max}]}$, with $\mathrm{t_{min}=-(1+\mu_c)^{1/\gamma}}$ and $\mathrm{t_{max}=(1-\mu_c)^{1/\gamma}}$. If $\mathrm{\tilde{\rho}(\lambda,t,\phi)>\chi}$ then the generated values of $\mathrm{\lambda}$, $\mathrm{t}$ and $\mathrm{\phi}$ are accepted as initial conditions, otherwise, they are rejected and the sampling process starts again.

Following the above described formulation, we obtain the initial conditions of the electron with respect to the origin of the coordinate system. To obtain the initial conditions for the position of the electron with respect to the center of mass of the triatomic molecule, $\mathrm{{\bf r'}}$,  in terms of the ones with respect to the origin, $\mathrm{{\bf r}}$,  we shift the coordinates by $\mathrm{{\bf r'}={\bf r}-{\bf R_{cm}}}$, where $\mathrm{{\bf R_{cm}}}$ is given by $\mathrm{(X_{cm},0,Z_{cm})}$ with
\begin{eqnarray}
X_{cm}&=&\frac{m_{C}x_{c}}{m_{A}+m_{B}+m_{C}},\nonumber\\
&&\\
Z_{cm}&=&\frac{R_{ab}(m_{B}-m_{A})/2+m_{C}z_{c}}{m_{A}+m_{B}+m_{C}}, \nonumber
\end{eqnarray}
with $\mathrm{m_{A}}$, $\mathrm{m_{B}}$ and $\mathrm{m_{C}}$ the masses of the nuclei.

 Next, using the one-electron microcanonical distribution we formulated above, we compute the position and momentum  probability densities  of the initially bound electron for $\mathrm{H_{3}^{+}}$. We do so for  the ground state of $\mathrm{H_{3}^{+}}$, where the nuclei are on the apexes of an equilateral triangle and the inter-nuclear distance is R=1.65 a.u. \cite{H3_GS_R_Schwartz1967}.  The first and second ionization energies are  $\mathrm{I_{p_1}}=1.2079$ a.u.  and $\mathrm{I_{p_2}}=1.93$ a.u., respectively, which we obtain  using MOLPRO, a quantum chemistry package   \cite{molpro2009}.  For the microcanonical distribution the relevant ionization energy is $\mathrm{I_p=I_{p_{2}}}$, since $\mathrm{I_{p_1}}$ is associated with the electron that tunnel-ionizes in the initial state.

\begin{figure} [ht!]
\centering
 \includegraphics[clip,width=0.6\textwidth]{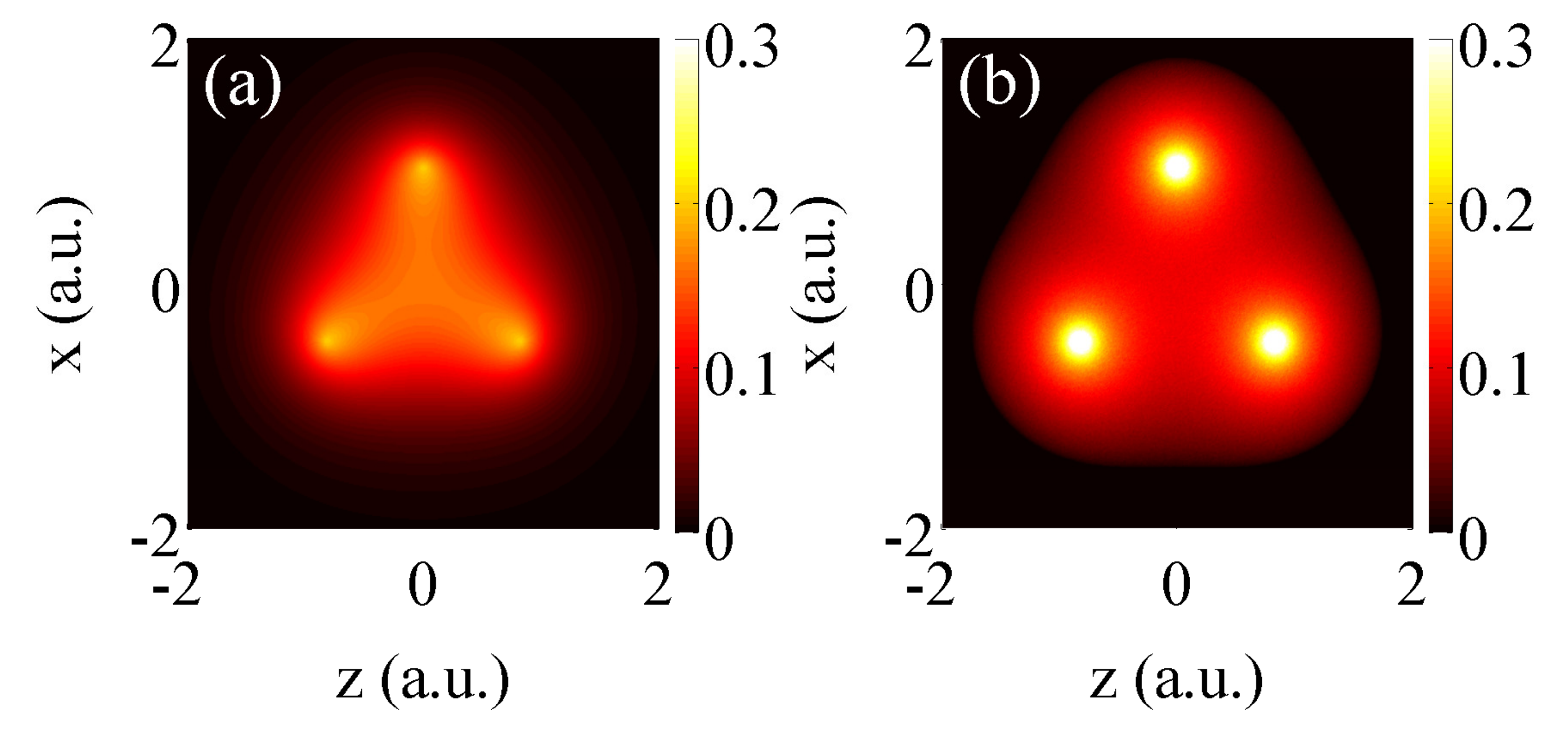}
\caption{(color online) Left panel: the quantum mechanical probability density of the electron position on the x-z plane for $\mathrm{y=0}$. Right panel: the  microcanonical probability density of the electron position on the x-z plane for $\mathrm{y=0}$.}
\label{position}
\end{figure}

In \fig{position} (b) we plot the probability density of the position of the electron on the x-z plane for $\mathrm{y=0}$. We compare this microcanonical distribution  with the quantum mechanical probability density of the position of the electron on the x-z plane in \fig{position} (a). That is, we plot $\mathrm{|\Psi(x,0,z)|^2}$, where $\mathrm{\Psi({\bf r})}$ is the quantum mechanical wavefunction for the $\mathrm{H_{3}^{2+}}$ molecule, which we obtain using MOLPRO. The two plots in \fig{position} show that the two probability densities of the electron position compare well. However, the microcanonical probability density underestimates the electron density  between the nuclei while it overestimates the one around the nuclei. 

\begin{figure} [ht!]
\centering
 \includegraphics[clip,width=0.8\textwidth]{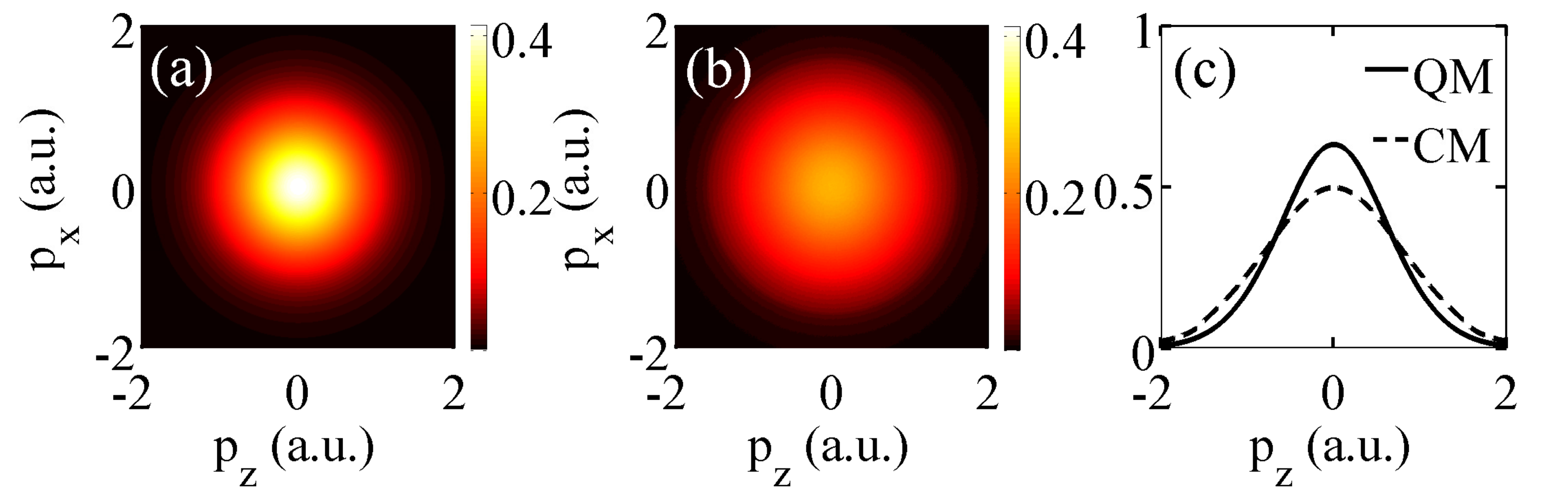}
\caption{(color online) Left panel: the quantum mechanical probability density of the electron momentum  on the $\mathrm{p_{x}-p_{z}}$ plane for all values of $\mathrm{p_{y}}$. Middle panel: the  microcanonical probability density of the electron momentum plotted on the $\mathrm{p_{x}-p_{z}}$ plane for all values of $\mathrm{p_{y}}$. Right panel: the projections on the $\mathrm{p_{z}}$ axis of the probability densities plotted in \fig{momentum} (a) and (b). }
\label{momentum}
\end{figure}

In addition, using the microcanonical distribution,  for all values of the electron momentum component along the y-axis, $\mathrm{p_{y}}$, we plot  the probability density  of the electron momentum on the $\mathrm{p_x-p_z}$ plane in \fig{momentum} (b). We compare this distribution with its quantum mechanical analog $\mathrm{\rho^{QM}(p_{x},p_{z})}$; the latter is plotted in  \fig{momentum} (a). To obtain $\mathrm{\rho^{QM}(p_{x},p_{z})}$, we, first, compute the quantum mechanical wavefunction in momentum space 
\begin{eqnarray}
\mathrm{\Phi({\bf{p}})}=\mathrm{\frac{1}{(2\pi)^{3/2}}\int\Psi({\bf{r}})e^{-i{\bf{pr}}}d{\bf{r}}},
\end{eqnarray}
and we, next, integrate  over $\mathrm{p_{y}}$
\begin{eqnarray}
\mathrm{\rho^{QM}(p_{x},p_{z})=\int_{-\infty}^{\infty}|\Phi({\bf p})|^2dp_{y}}.
\end{eqnarray}
The two plots in \fig{momentum}  show that the two probability densities for the electron momentum compare well. However, the microcanonical probability density overestimates the higher values of the electron momentum. This can be clearly seen in \fig{momentum} (c) where we project the probability densities of the electron momentum plotted in  \fig{momentum} (a) and (b) on the $\mathrm{p_{z}}$ axis. The higher values of the electron momentum resulting from the microcanonical distribution  are  consistent with our finding that the microcanonical distribution overestimates the values of the electron position around the nuclei.

\section{Results}

In what follows, we investigate three different processes that take place through Coulomb explosion during  the fragmentation of $\mathrm{H_{3}^{+}}$, when the molecule is driven by a near-IR intense laser field. 
Specifically, we study: i) double ionization where the final fragments are three $\mathrm{H^{+}}$ ions and two escaping electrons; ii) frustrated double ionization where the final fragments are a highly excited neutral fragment $\mathrm{H^{*}}$,  two $\mathrm{H^{+}}$ ions and one escaping electron; iii) frustrated single ionization  where the final fragments are  two highly excited neutral fragments $\mathrm{H^{*}}$ and one $\mathrm{H^{+}}$ ion. We mainly focus on FDI and FSI. 

\subsection{The model}
The laser field we use in our model is of the form   
\begin{eqnarray}
\begin{split}
&\mathrm{{\bf{E}}(t)=E_{0}(t)[\cos({\omega t})\hat{z}+\epsilon\sin({\omega t}) \hat{x}]}\\
&\mathrm{E_{0}(t)}=\left\{ 
\label{eqnfield}
\begin{array}{lcc}
\mathrm{E_{0}} && \mathrm{0 \leq t < 10T}\\
\mathrm{E_{0}\cos^2\frac{\omega(t-10T )}{8}} && \mathrm{10T \leq t \leq 12T},
\end{array}
\right.
\end{split}
\end{eqnarray}
where $\omega= 0.057$ a.u. (800 nm),  $\epsilon$, $\mathrm{E_{0}(t)}$  and $\mathrm{T}$ are the frequency, the ellipticity, the envelope and the period of the laser field, respectively. Atomic units are used throughout this work unless otherwise indicated.  

We take the initial state to be the ground state  of $\mathrm{H_{3}^{+}}$ with the nuclei forming an equilateral triangle with an inter-nuclear distance  R=1.65 a.u. \cite{H3_GS_R_Schwartz1967}. In our simulations,  we take the three nuclei A, B and C to be on the x-z plane. In addition, we simplify our model by considering the nuclei initially at rest since an initial pre-dissociation does not significantly modify the ionization dynamics \cite{Toolkit_Agapi2014}. We note that for $\mathrm{I_{p_1}=1.208}$ a.u. we find that the threshold field strength $\mathrm{E_{0}}$ for over-the-barrier ionization is 0.178 a.u.. If the instantaneous field strength at the time we start the propagation is smaller than the threshold field strength for over-the-barrier ionization, we assume that one electron (electron 1) tunnels in the field-lowered Coulomb potential with a tunneling rate given by the semi-classical formula in \cite{ionization_rate_Murray2011}. The tunnel electron emerges from the potential barrier with zero velocity along the direction of the laser field and with a velocity that follows a  Gaussian distribution in the direction perpendicular to the laser field \cite{Perpendicular_p_Delone1991}.  If the instantaneous field strength at the time we start the propagation corresponds to the over-the-barrier intensity regime, then we assume that electron 1 tunnel ionizes at the maximum of the field lowered Coulomb potential. We take  the kinetic  energy of electron 1 to be  equal to the difference between the first ionization energy and the maximum of the field-lowered Coulomb potential, for details see \cite{Toolkit_Agapi2014}. For both below- and over-the-barrier ionization of electron 1 in the initial state, we describe the initial state of the initially  bound electron (electron 2) using the one-electron microcanonical distribution which we presented in section II.

\subsection{FDI and FSI  in  $\mathrm{H_{3}^{+}}$ when driven by a linearly polarized field}
\begin{figure} [ht]
\centering
  \includegraphics[clip,height=0.25\textwidth]{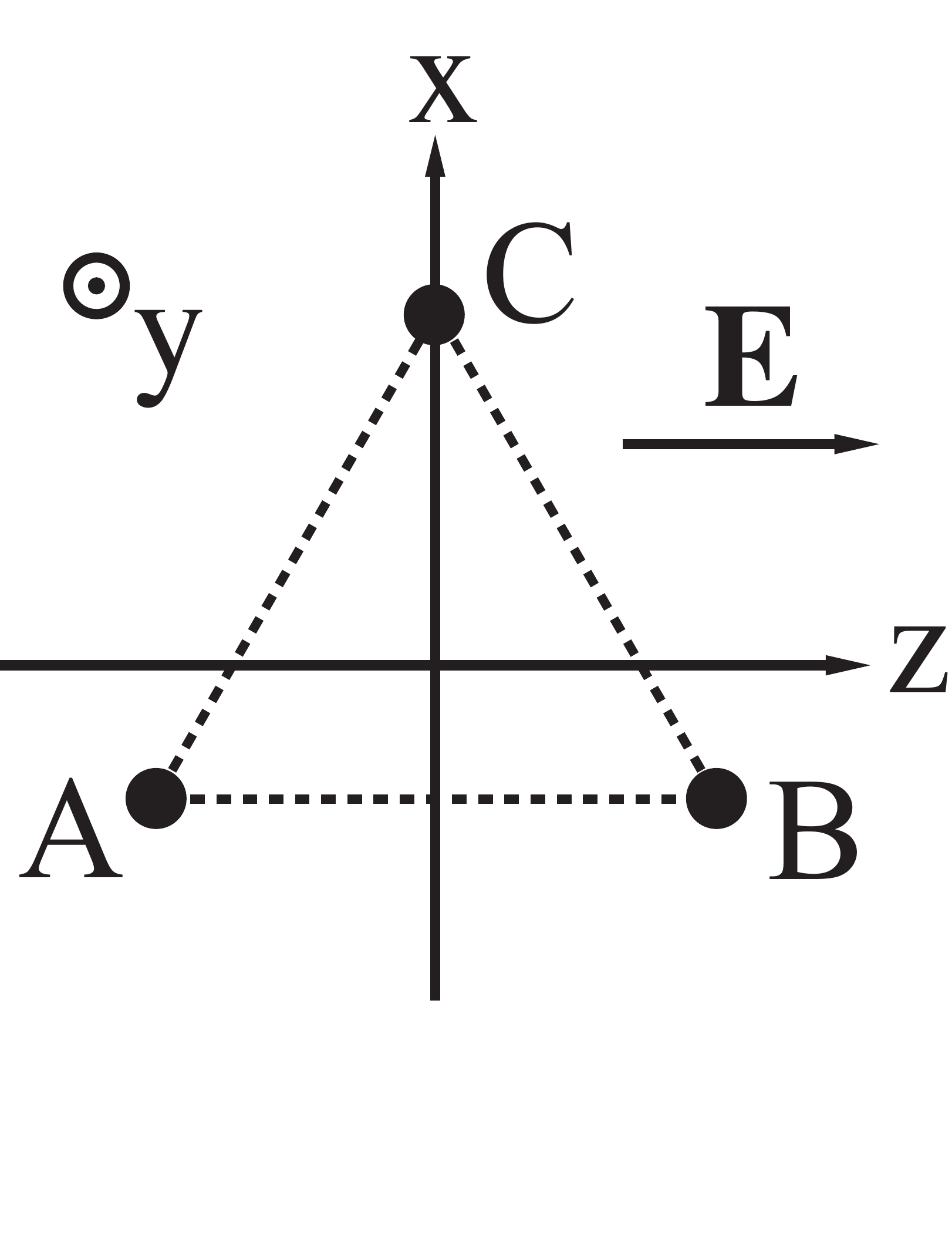}
  \hspace{0.15in}
  \includegraphics[clip,height=0.25\textwidth]{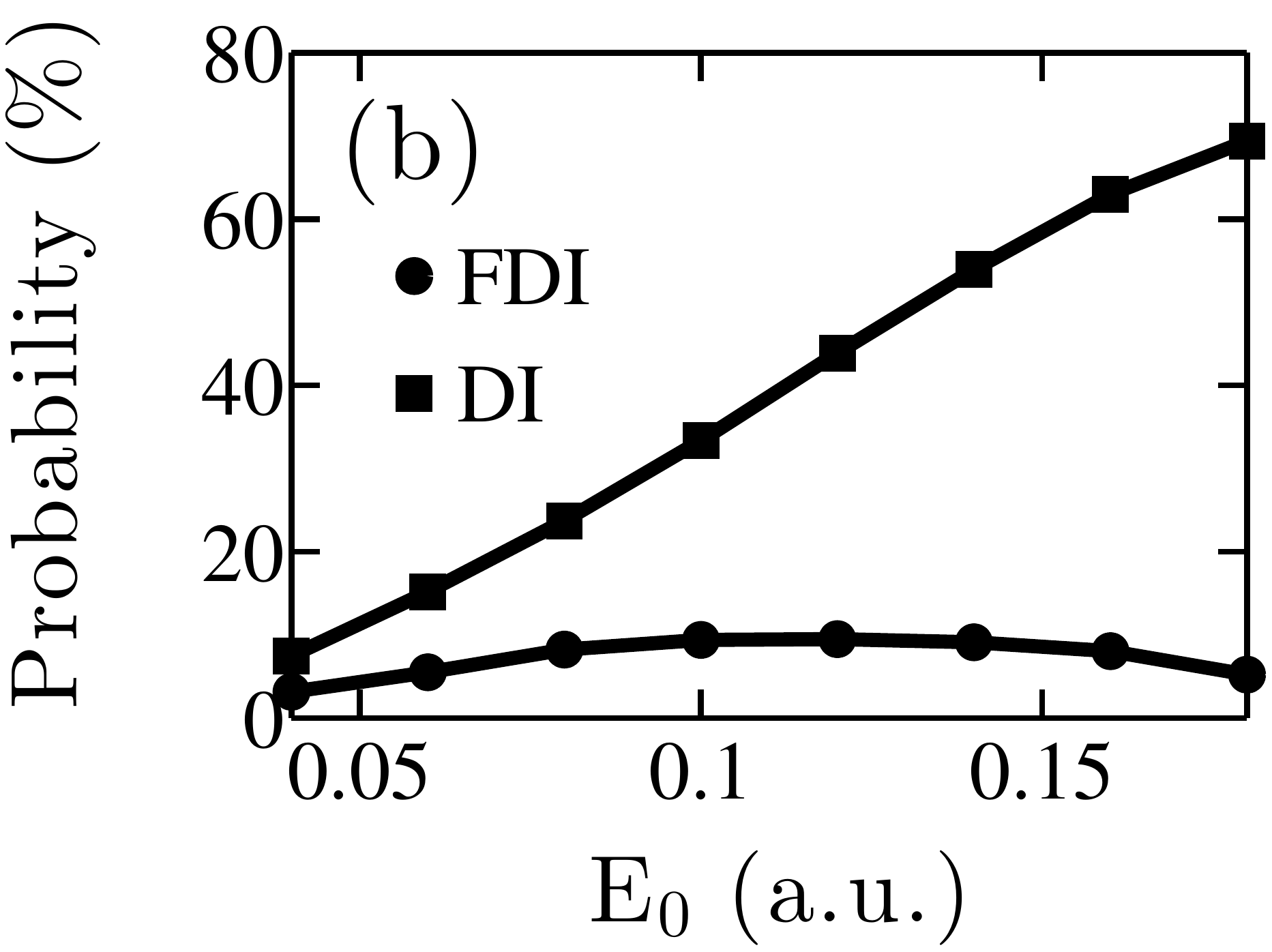}
\caption{Left panel: The initial configuration of the nuclei in the  $\mathrm{H_{3}^{+}}$ triatomic molecule. The laser pulse is linearly polarized and aligned along one side of the triatomic molecule.
Right panel: The  DI and FDI  probabilities as a function of the laser field strength $\mathrm{E_0}$. The lowest laser field strength in (b) is E$_{0}=0.04$ a.u..}
\label{fig1}
\end{figure}

 We consider a laser field  linearly polarized along one side of the equilateral triangle, see \fig{fig1} (a). In \fig{fig1} (b), we plot the DI and FDI  probabilities as a function of the laser field strength. We vary the laser field strength from  0.04 a.u. up to 0.18 a.u., that is up to a field strength   just above the threshold value for over-the-barrier ionization. In this context, probability is the number of DI, FDI and FSI events relative to the number of initialized trajectories. We find that DI is the dominant process at E$_0$=0.18 a.u. with a probability of 69.4\%. The FDI probability reaches a maximum of 9.5\% at E$_{0}=0.12$ a.u. and reduces to 5.2\% at E$_{0}= 0.18$ a.u..

Focusing on FDI during the fragmentation  of strongly-driven $\mathrm{H_{3}^{+}}$ from its ground state, we find that two main pathways, A and B, contribute to FDI. We have previously identified these two pathways in our studies of FDI during the  fragmentation of strongly-driven H$_{2}$ from its ground state \cite{semiclassical_agapi2012} and of strongly driven D$_{3}^{+}$ from a state other than its  ground state \cite{Agapi2016submitted}. As in our previous studies, we find that in pathway A, electron 1 escapes, while  electron 2 tunnel-ionizes later while the field is on and is eventually recaptured to a highly excited state of an H atom (H$^*$). In pathway B, electron 1   is eventually recaptured to a highly excited state of H, while electron 2 tunnel-ionizes later but eventually escapes. In pathway A, electron 2 tunnels after gaining energy in a frustrated enhanced ionization process, i.e., it gains energy from the field in the same way as in an enhanced ionization process \cite{Enhanced_Niikura2002,Enhanced_Bandrauk1995,enhanced_Seideman1995,enhanced_Villeneuve1996,enhanced_Dehghanian2010,enhanced_wu2012} but electron 2 eventually does not escape. Electron-electron correlation is more important for pathway B than for pathway A. This is to be expected since in pathway B electron 1, following tunnel-ionization, returns to the core and interacts with electron 2.  

\begin{figure} [ht]
\centering
  \includegraphics[clip,height=0.3\textwidth]{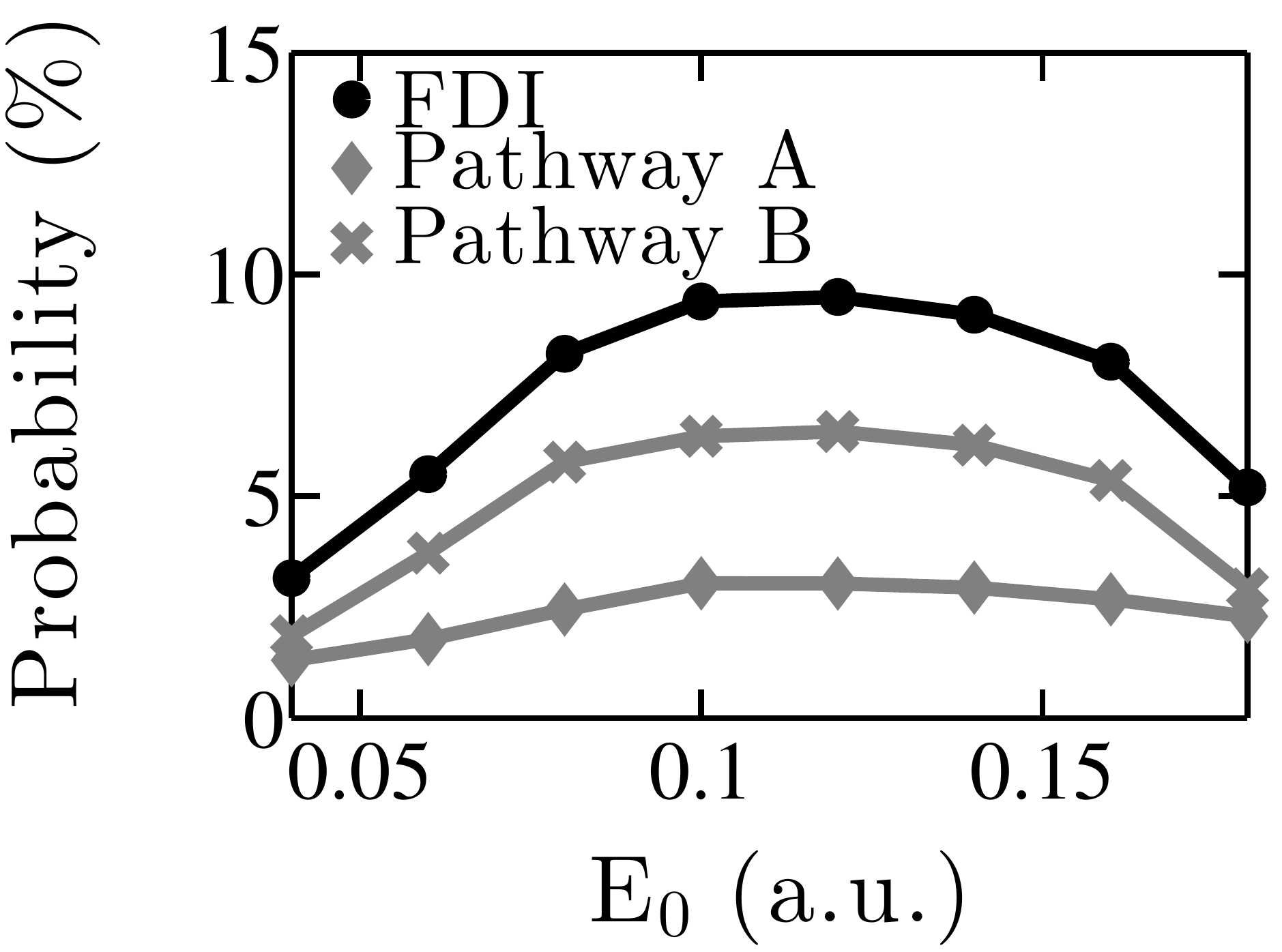}
\caption{The  FDI probability   and the probabilities for pathways A and B of FDI as a function of the laser field strength $\mathrm{E_0}$. The lowest laser field strength  is E$_{0}=0.04$ a.u..
}
\label{fig_PAPB}
\end{figure}

 \fig{fig_PAPB} shows that  for intermediate strengths of the laser field  below the over-the-barrier ionization threshold,   pathway B is the dominant pathway of FDI. \fig{fig_PAPB} also shows that
pathway A's  contribution  to FDI increases with increasing field strength. At E$_{0}=0.18$ a.u. both pathways have the same probability. These results are not surprising since, in strongly-driven molecules,  electron-electron correlation is more important for intermediate strengths of the laser field, while enhanced ionization becomes more prominent  with increasing  strength of the laser field.

\begin{figure} [ht]
\centering
  \includegraphics[clip,height=0.3\textwidth]{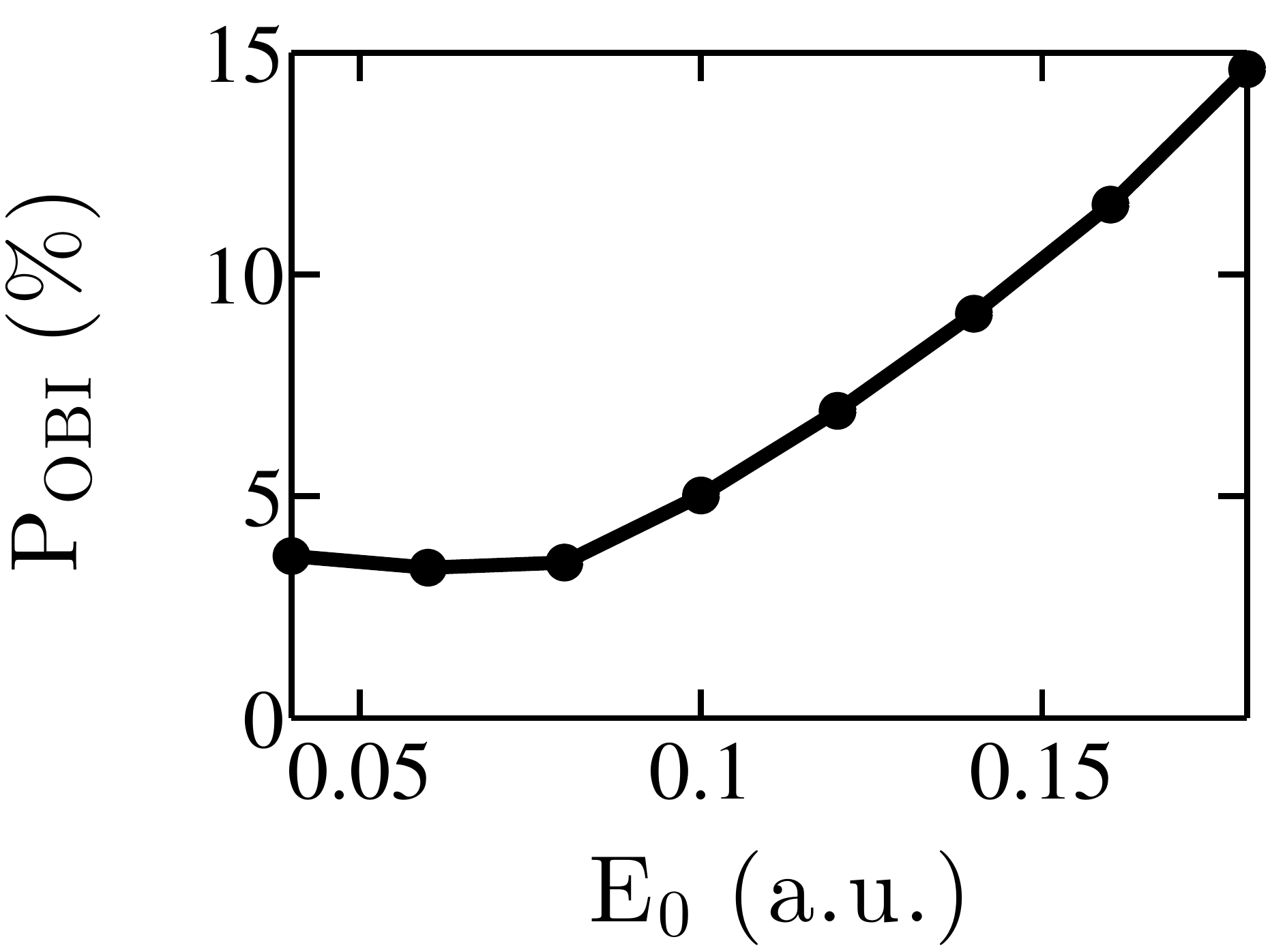}
\caption{ $\mathrm{P_{OBI}}$ as a function of the laser field strength.  The lowest laser field strength  is E$_{0}=0.04$ a.u..}
\label{fig_OBI}
\end{figure}

Next, we  investigate  whether tunnel-ionization is the underlying mechanism of FDI, as it was first suggested in \cite{FTI_exp_Nubbemeyer2008}. Specifically, we check whether tunnel or over-the-barrier ionization is the underlying mechanism of FDI.  By over-the-barrier ionization in FDI, we refer to electron 2    reaching an excited state without tunneling in pathway A or  to electron 2 escaping by over-the-barrier ionization in pathway B. We denote by $\mathrm{P_{OBI}}$ the fraction of FDI over-the-barrier ionization events   out of all FDI events. As shown in \fig{fig_OBI}, $\mathrm{P_{OBI}}$ increases from 3.7\% at E$_0$=0.04 a.u. to 14.6\% for E$_0$=0.18 a.u. This increase of $\mathrm{P_{OBI}}$ is due to over-the-barrier ionization becoming more prominent with increasing strength of the laser field. Very recently, we obtained similar results for the contribution of the over-the-barrier ionization mechanism in FDI for D$_3^{+}$ when this molecule is strongly-driven from an initial state created via the reaction 
$\mathrm{D_{2}+D_{2}^{+}\rightarrow D_{3}^{+}+D}$ \cite{Agapi2016submitted}.

\begin{figure}[ht!]
\centering
 \includegraphics[clip,height=0.25\textwidth]{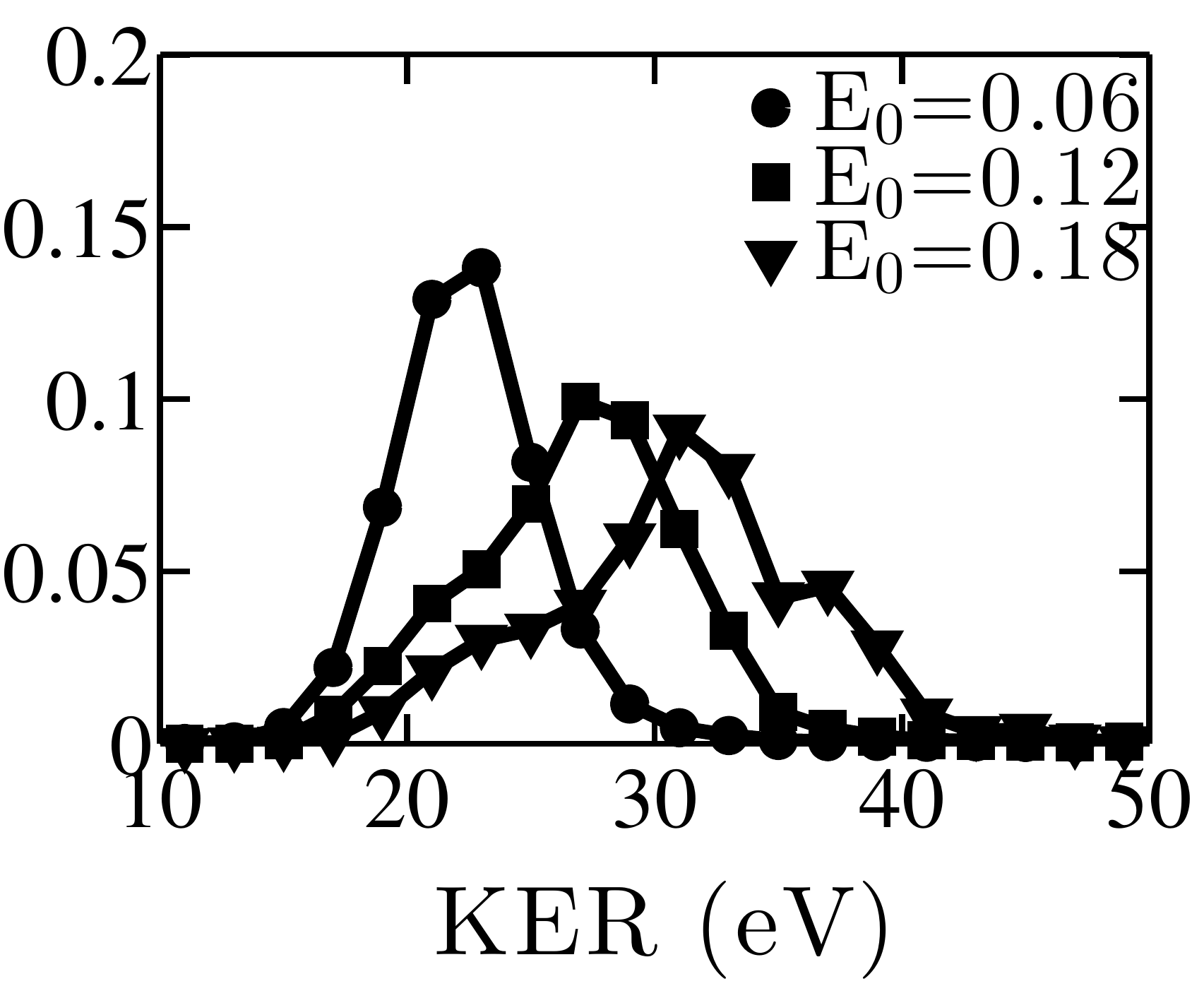}
 \caption{ The KER distributions   for FDI  at laser field strengths of 0.06 a.u., 0.12 a.u. and 0.18 a.u.. } 
\label{figKERFDI}
\end{figure}

 The kinetic energy release (KER), that is, the sum of the kinetic energies of the final ion fragments is a quantity often measured in experiments  \cite{D3+_exp_McKenna2012}. In \fig{figKERFDI}, we plot the KER distribution  for FDI for three different laser field strengths. We find that with increasing strength of the laser field the peak of the  KER distribution shifts to higher values, namely, from 23 eV at  $\mathrm{E_0}=$0.06 a.u.  to 31 eV at  $\mathrm{E_0}=$0.18 a.u.. This increase is consistent  
with the nuclei Coulomb exploding earlier in time and at smaller inter-nuclear distances for higher strengths of the laser field. 

\begin{figure}[ht!]
\centering
  \includegraphics[clip,height=0.25\textwidth]{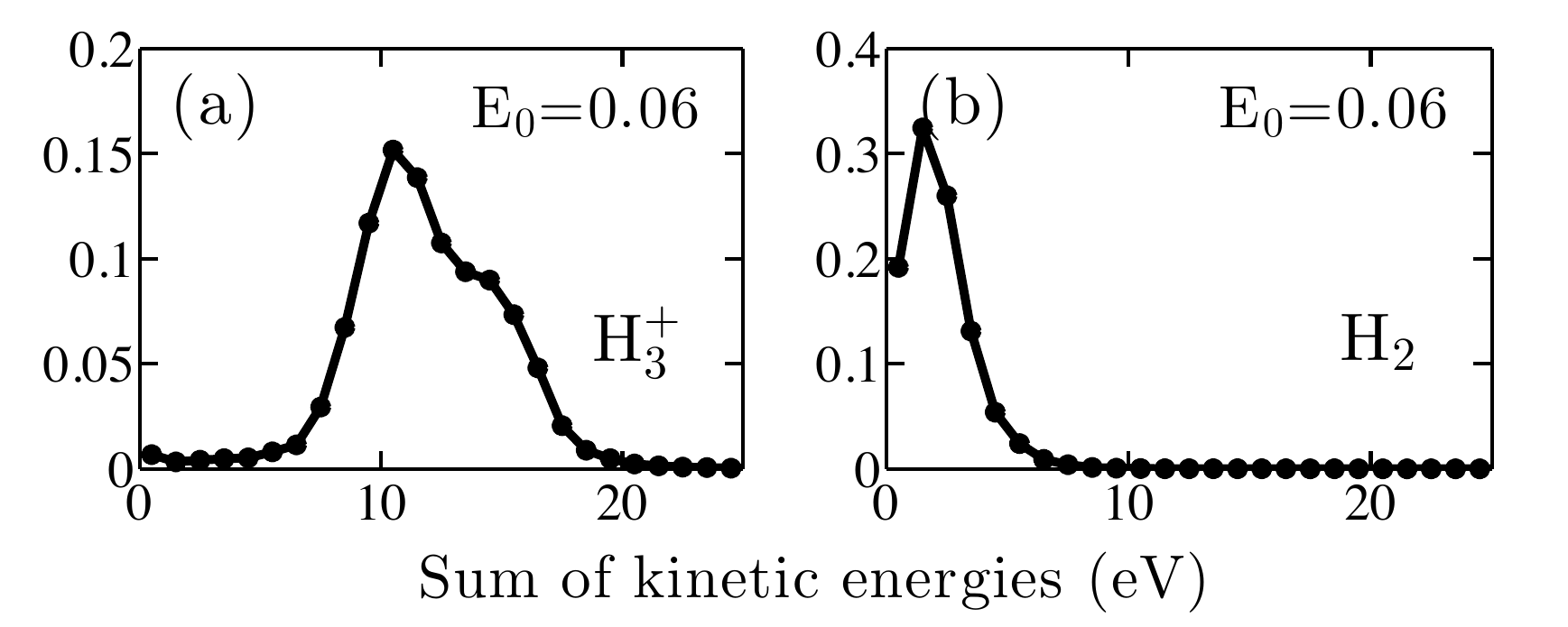}
\caption{
The distribution of the sum of the kinetic energies of the nuclei   for FDI at the time electron 2 tunnels   for (a) $\mathrm{H_3^+}$  and for (b) $\mathrm{H_2}$ at $\mathrm{E_0}=$0.06 a.u.. } 
\label{figKERinitial}
\end{figure}

Previously, both for the fragmentation of strongly-driven H$_{2}$ from its ground state \cite{Toolkit_Agapi2014}  and of strongly driven D$_{3}^{+}$ from a superposition of states with inter-nuclear distances larger than the inter-nuclear distance of the ground state \cite{Agapi2016submitted}, we have found that the peak of the KER distributions can be roughly estimated as follows. We first compute the most probable distance  of the nuclei at the time electron 2 tunnels, R$_\mathrm{{tun}}$. For the above-mentioned previous studies,  this is also the time when Coulomb explosion of the nuclei mostly sets in. As a result, we found that the KER distributions peak roughly at 2/R$_\mathrm{{tun}}$ for H$_{2}$ and at 3/R$_\mathrm{{tun}}$ for D$_{3}^{+}$. We find that this is not  quite the case for strongly driven H$_{3}^{+}$ when driven from its ground state.   
Specifically,  in \fig{figKERinitial}, we plot the sum of the kinetic energies of the ions  at the time electron 2 tunnels  for strongly-driven H$_3^{+}$ (a) and H$_{2}$ (b) at E$_{0}=0.06$ a.u.. We show that the distribution of the sum of the kinetic energies of the nuclei 
for H$_3^{+}$ peaks around 10.5 eV, see \fig{figKERinitial} (a), while  for H$_2$ the distribution peaks around 1.5 eV, see \fig{figKERinitial} (b). Thus, for H$_{3}^{+}$ the nuclei have already acquired a significant amount of kinetic energy by the time electron 2 tunnels unlike H$_{2}^{+}$. This is reasonable since one electron screens more effectively two rather than three nuclei.  For H$_{3}^{+}$ fragmenting from its ground state, to roughly estimate where the KER distribution for FDI peaks we have to add 3/R$_\mathrm{tun}$+10.5= 25.3 eV; we have substituted  $\mathrm{R_{tun}=5.5}$ a.u. which we obtain from our simulations. Indeed, we find that  the KER distribution for FDI peaks at 23 eV, see \fig{figKERFDI}, which is slightly less than 25.3 eV, since for FDI  the electron that is recaptured screens  the Coulomb explosion of the nuclei.

\begin{figure}[ht!]
\centering
\includegraphics[clip,height=0.25\textwidth]{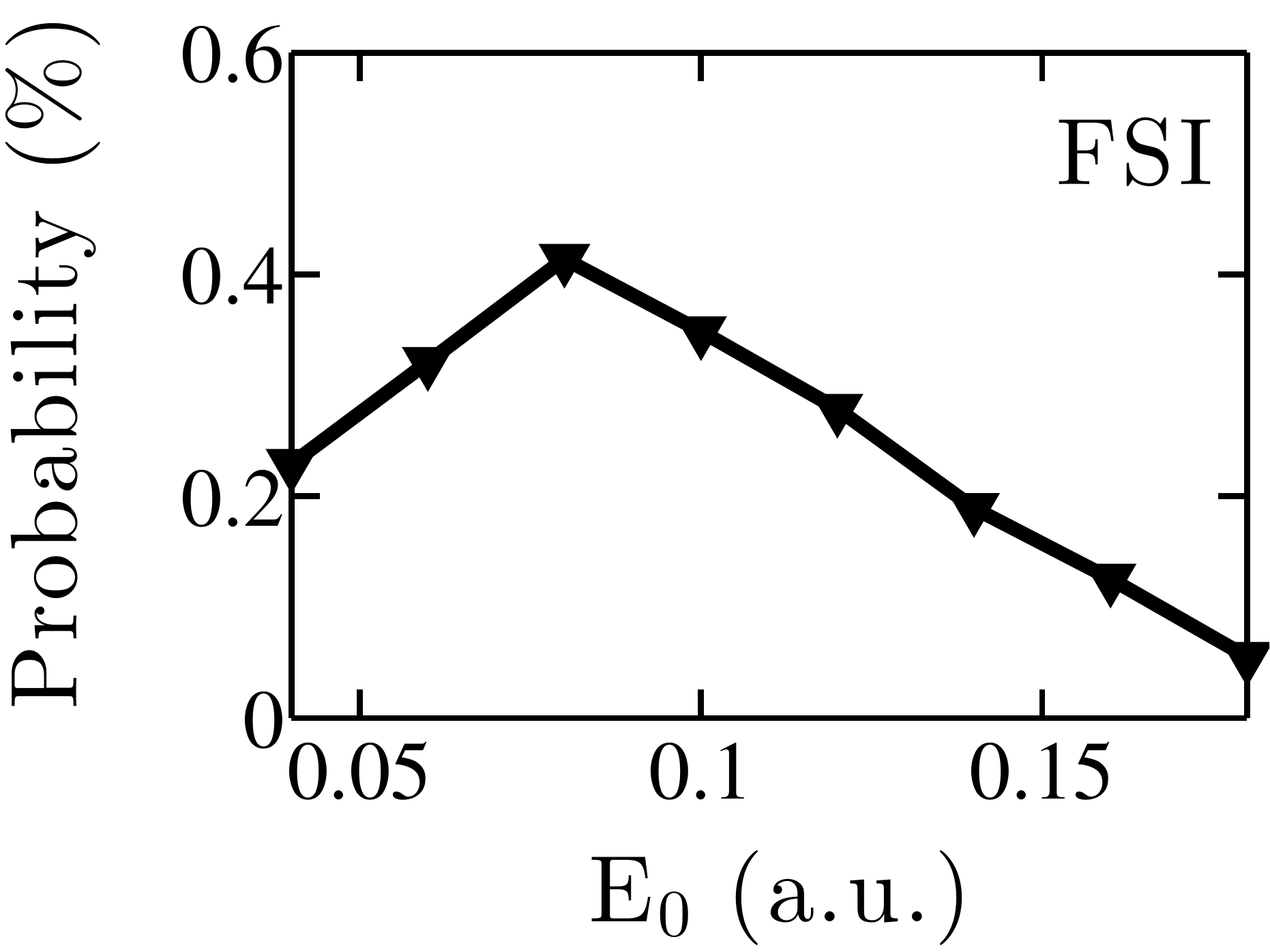}
\caption{FSI probability as a function of the laser field strength. The lowest laser field strength  is E$_{0}=0.04$ a.u..} 
\label{figFSIprob}
\end{figure}

\begin{figure}[ht!]
\centering
 \includegraphics[clip,height=0.25\textwidth]{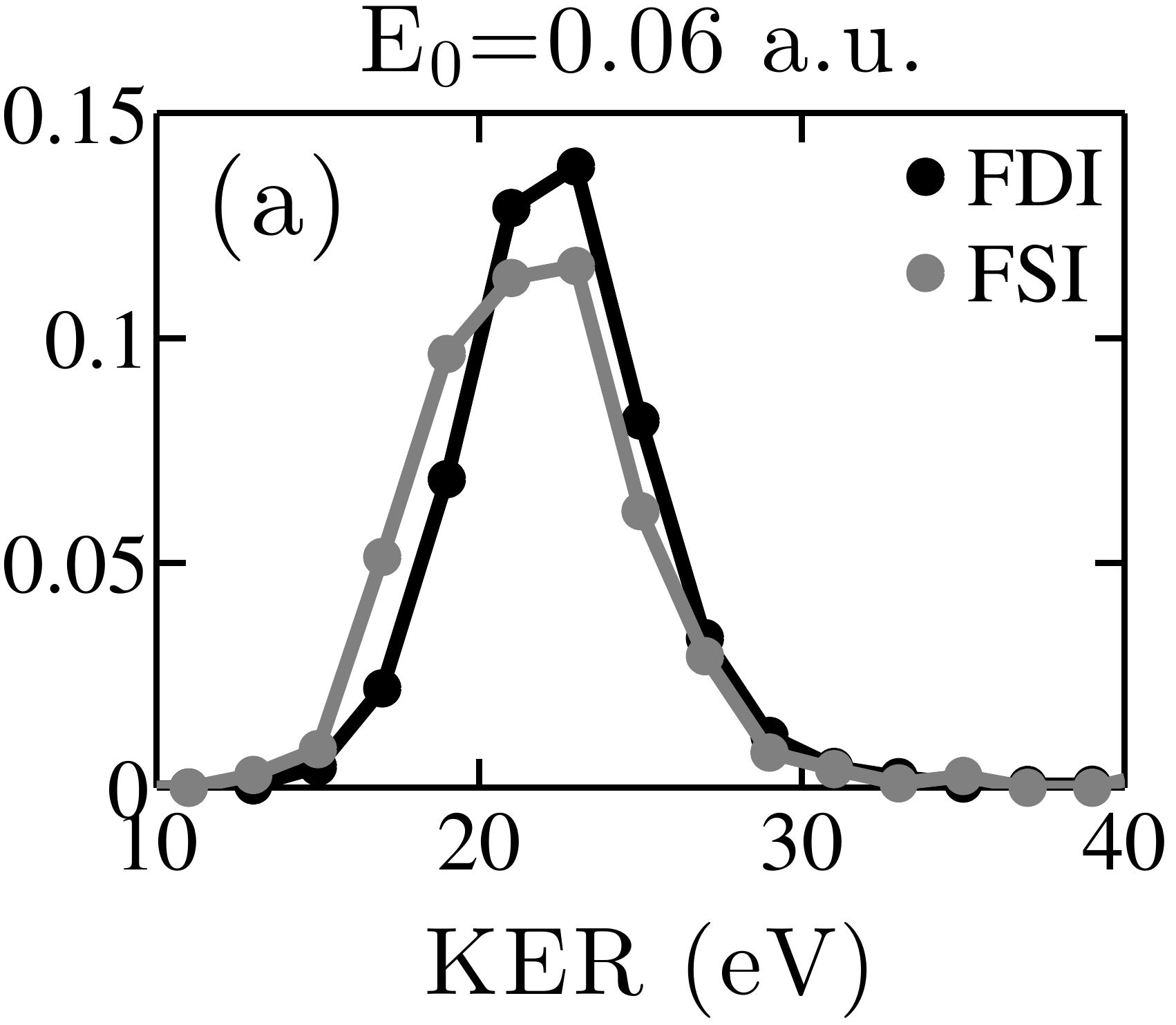}
 \includegraphics[clip,height=0.25\textwidth]{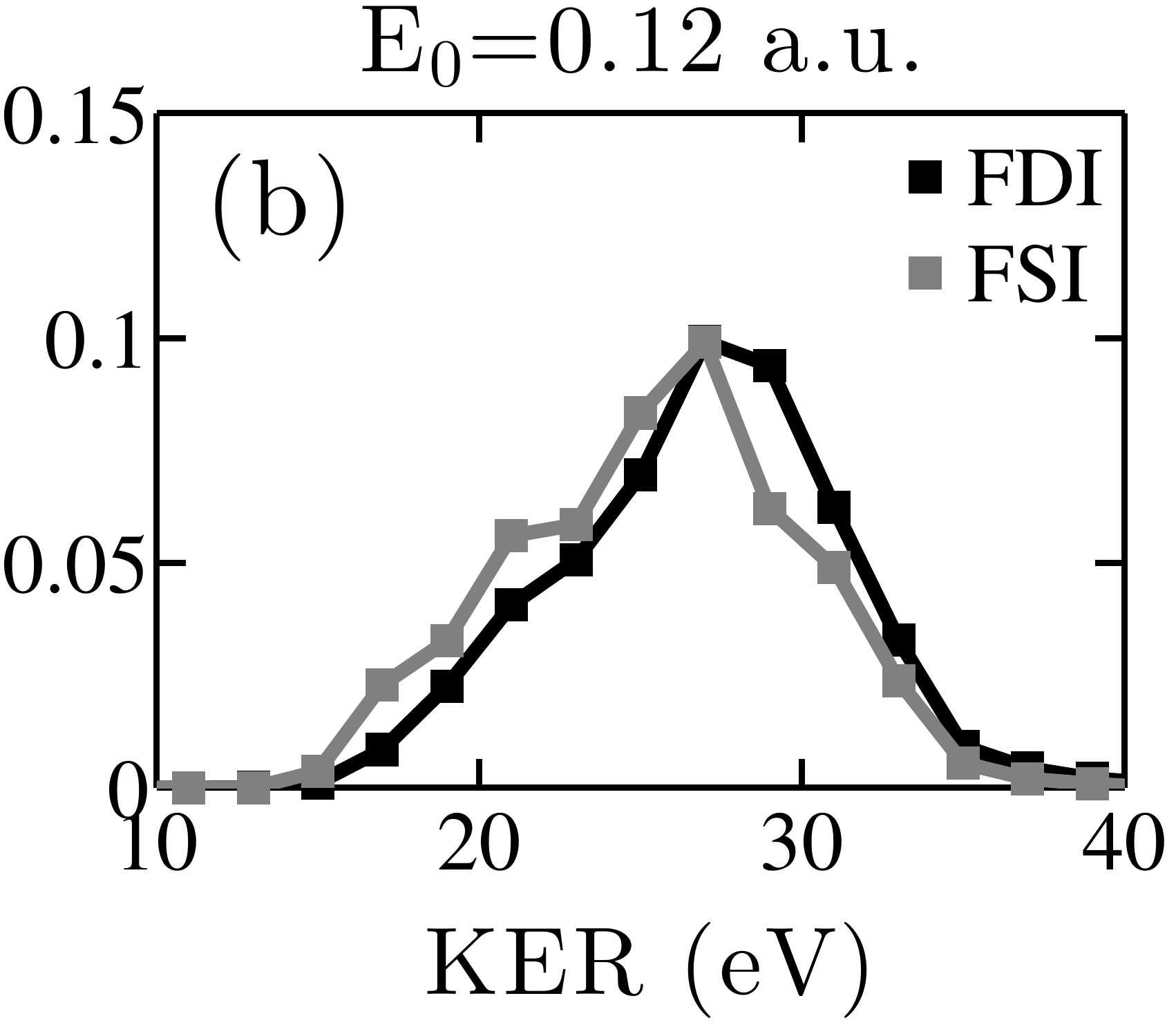}
\caption{KER distributions for FSI at laser field strengths of 0.06 a.u. and 0.12 a.u.. The KER distributions for FDI are also presented for comparison.} 
\label{FSI_KER}
\end{figure}

 Next, we address FSI where two highly excited neutrals are formed:
 \begin{equation}
 \mathrm{H}_3^+ \rightarrow \mathrm{H}^{\star} + \mathrm{H}^{\star} + \mathrm{H}^+. \nonumber
\end{equation}
 In \fig{figFSIprob}, we show that the FSI probability reaches a maximum probability of 0.4\% at E$_{0}=0.08$ a.u. and then reduces fast with increasing field strength reaching 0.06\% at E$_{0}=0.18$ a.u.. This is consistent with a higher strength of the laser field resulting in a higher probability for an electron to ionize. Thus, FSI is a process roughly 20 times less likely than FDI.
 We have reached a similar conclusion in previous studies of FSI in the context of strongly-driven H$_{2}$ \cite{Agapi2012NJP}. As for FDI, we plot the KER distributions for FSI  for different laser field strengths, see \fig{FSI_KER}.  We find that the KER distributions for FSI  peak at similar energy values as the KER distributions for FDI.
 
\subsection{Influence of molecular geometry on FDI }
In what follows, we investigate whether a different molecular geometry affects  the FDI probability. We do so, by comparing our results for  FDI  for the diatomic $\mathrm{H_{2}}$ with the triatomic $\mathrm{H_{3}^{+}}$.
First, we consider that both molecules  are driven by a linearly polarized laser field of the same laser field strength.  The laser field is aligned with one side of the molecules. In Table II, we show the results for laser field strengths of  0.04 a.u. and 0.06 a.u.. We find that the FDI probability  is much larger for H$_{2}$. This result is not surprising. It is easier to ionize an electron in the diatomic molecule, since  both molecules are driven with the same laser field strength while the ionization energies of H$_{3}^{+}$ are much larger than those of  H$_{2}$. The first and second ionization energies of $\mathrm{H_{3}^{+}}$ are $\mathrm{I_{p_1}=}$1.2079 a.u.  and $\mathrm{I_{p_2}=}$1.9300 a.u., respectively, while for  $\mathrm{H_{2}}$  $\mathrm{I_{p_1}=}$0.5669 a.u.  and $\mathrm{I_{p_2}=}$1.2843 a.u.

Next, we compare  the FDI probability  when the ionization  probability $\mathrm{\Gamma(I)}$ is the same. 
The ionization probability is obtained by integrating, over the duration of  the laser pulse, the ionization rate $\mathrm{\Gamma( t,I)}$ for a  laser pulse  intensity I: 
\begin{eqnarray}
\mathrm{\Gamma(I)}\approx\int^{t_f}_{t_i}\mathrm{\Gamma(t, I)}dt.
\end{eqnarray}
In Table III, we present the FDI probability  for $\mathrm{H_{2}}$ and  $\mathrm{H_{3}^{+}}$ when $\mathrm{\Gamma(I)}$ is of the order of 10$^{-5}$ and 10$^{-2}$. We find that  when $\mathrm{\Gamma(I)}$ is the same for both molecules the FDI probability is also roughly the same.
Moreover, we find that the probability for pathway B of FDI is for both molecules larger than the probability for pathway A of FDI.  The above results suggest that the molecular geometry does not significantly affect the FDI probability.

\begin{table} [t!] 
\begin{center}
\footnotesize
\begin{tabular}[t]{|c|c|c|c|c|c|c|c|c|}
\hline
Molecule&$\mathrm{E_0} (a.u.)$ &$\mathrm{\Gamma(I)}$&FDI (\%)& Pathway A (\%)&Pathway B (\%)&DI (\%) \\
\hline
H$_2$&0.04& $4.0\times10^{-5}$&9.6& 2.9 & 6.7 &24.4\\
\hline
H$^+_3$&0.04&$2.0\times10^{-21}$&3.2&1.3  & 1.9&7.5\\
\hline
H$_2$&0.06&0.03&9.4& 3.2& 6.1&39.5\\
\hline
H$^+_3$&0.06 &$1.5\times10^{-12}$&5.5&1.8  & 3.7&15.2\\
\hline
\end{tabular}\\
\caption{The FDI and DI probabilities and the probabilities for pathways A and B of FDI  for $\mathrm{H_{2}}$ and $\mathrm{H_{3}^{+}}$ in a linearly polarized laser field for laser field strengths of 0.04 a.u. and 0.06 a.u..}
\end{center}
\end{table}

\begin{table} [t!] 
\begin{center}
\footnotesize
\begin{tabular}[t]{|c|c|c|c|c|c|c|c|c|}
\hline
Molecule&$\mathrm{E_0} (a.u.)$ &$\mathrm{\Gamma(I)}$&FDI (\%)& Pathway A (\%)&Pathway B (\%)& DI (\%) \\
\hline
H$_2$&0.04& $4.0\times10^{-5}$&9.6& 2.9 & 6.7 &24.4\\
\hline
H$^+_3$&0.10&$1.5 \times10^{-5}$&9.4& 3.0 & 6.3&33.3\\
\hline
H$_2$&0.06&0.03&9.4& 3.2& 6.1&39.5\\
\hline
H$^+_3$&0.15&0.04&8.7  & 2.8&5.8&58.3\\
\hline
\end{tabular}\\
\caption{The FDI and DI probabilities and the probabilities   for pathways A and B of FDI  for $\mathrm{H_{2}}$ and $\mathrm{H_{3}^{+}}$ in a linearly polarized laser field. The laser field strengths are chosen so that the two molecules have similar ionization probabilities $\mathrm{\Gamma(I)}$.}  \label{table_h2h3}
\end{center}  
\end{table}

\section{Conclusions}
We formulate a microcanonical distribution for arbitrary one-electron triatomic molecules. In the current work, we use this microcanonical distribution to describe the initial state of the bound electron in our study of strongly-driven H$_{3}^{+}$ from its ground state. We show that the kinetic energy release distribution of the nuclei for FDI peaks at a higher energy than the one roughly estimated  from the Coulomb explosion of the nuclei at the time the bound electron tunnel-ionizes. The reason is that by the time the bound electron tunnel-ionizes the nuclei have already acquired a significant amount of kinetic energy. As we show, this is unlike the case of H$_{2}^{+}$ fragmenting from its ground state. In addition, we show that FSI is a more rare process compared to FDI which is a significant process with  probability 10\%. Finally, we show that the FDI probability is not significantly influenced by the different molecular geometry of H$_2$ and H$^+_3$.

{\it Acknowledgments.} A.E. acknowledges  the use of the computational resources of Legion at UCL and fruitful discussions with Andre Staudte.

\begin{flushleft}
\small
\bibliography{cite}
\end{flushleft}

\end{document}